\documentclass[aps,showpacs]{revtex4}
\usepackage{graphicx}
\usepackage{amssymb}
\usepackage{amsmath}
\newcommand{\be}{\begin{equation}}
\newcommand{\ee}{\end{equation}}
\newcommand{\ben}{\begin{eqnarray}}
\newcommand{\een}{\end{eqnarray}}
\newcommand{\bes}{\begin{subequations}}
\newcommand{\ees}{\end{subequations}}

\newcommand\lsim{\mathrel{\rlap{\lower4pt\hbox{\hskip1pt$\sim$}}
    \raise1pt\hbox{$<$}}}
\newcommand\gsim{\mathrel{\rlap{\lower4pt\hbox{\hskip1pt$\sim$}}
    \raise1pt\hbox{$>$}}}
\newcommand\esim{\mathrel{\rlap{\raise2pt\hbox{\hskip0pt$\sim$}}
    \lower1pt\hbox{$-$}}}
\begin{document}
\title{Quintessence and tachyon dark energy models with a constant equation of state parameter}
\author{P.P. Avelino}
\affiliation{Centro de F\'{\i}sica do Porto, Rua do Campo Alegre 687, 4169-007 Porto, Portugal}
\affiliation{Departamento de F\'{\i}sica da Faculdade de Ci\^encias
da Universidade do Porto, Rua do Campo Alegre 687, 4169-007 Porto, Portugal}
\author{L. Losano}
\affiliation{Departamento de F\'{\i}sica, Universidade Federal da Para\'{\i}ba 58051-970 Jo\~ao Pessoa, Para\'{\i}ba, Brasil}
\author{J.J. Rodrigues}
\affiliation{Departamento de F\'{\i}sica, Universidade Federal da Para\'{\i}ba 58051-970 Jo\~ao Pessoa, Para\'{\i}ba, Brasil}
\date{\today}

\begin{abstract}
In this work we determine the correspondence between quintessence and tachyon dark energy models with a constant dark energy equation of state parameter, $w_e$.  Although the evolution of both the Hubble parameter and the scalar field potential with redshift is the same, we show that the evolution of quintessence/tachyon scalar fields with redshift is, in general, very different. We explicity demonstrate that if $w_e \neq -1$ the potentials need to be very fine-tuned for the relative perturbation on the equation of state parameter, $\Delta w_e/(1+w_e) \ll 1$, to be very small around the present time. We also discuss possible implications  of our results for the reconstruction of the evolution of $w_e$ with redshift using varying couplings. 

\end{abstract}


\maketitle

%%%%%%%%%%%%%%%%%%%%%%%%%%%%%

\section{Introduction}

There is now overwhelming evidence for a recent acceleration of the expansion of the universe \cite{Frieman:2008sn,Komatsu:2010fb}. At present the observational data appears to be consistent with a constant dark energy density, also known as a cosmological constant, with a constant dark energy equation of state parameter, $w_e=-1$. However, dynamical dark energy is probably a more reasonable explanation for the observed acceleration of the expansion of the universe, taking into account the enormous discrepancy between observationally inferred vacuum energy density and theoretical expectations. Dynamical dark energy candidates include minimally coupled scalar fields, vector fields or even modifications to General Relativity on cosmological scales  \cite{Carroll:2004de,Copeland:2006wr}, such as those associated with extra-dimensions \cite{Dvali:2000hr,Avelino:2001qh,Afshordi:2008rd} or $f(R)$ theories \cite{Capozziello:2006dj,Nojiri:2006gh,Hu:2007nk,Sotiriou:2008rp}. 

Although current observations seem to be consistent with a constant $w_e=-1$, it is interesting to ask how likely it is for the dark energy parameter to be a constant other than $-1$. This question has been addressed in \cite{Avelino:2009ze} where it has been  shown that a considerable amount of fine-tuning of the quintessence scalar field potential would be required in order to obtain a constant $w_e \neq -1$. It was argued that if future evidence excludes the cosmological constant  as a dark energy candidate, that should be interpreted as very strong evidence in favor of dynamical dark energy, even if the data appears to be consistent with a time-independent value for $w_e$.

In this paper we revisit this problem in a broader context. We extend the correspondence between quintessence and tachyon models which has been extensively studied in \cite{2002PhRvD..66b1301P,2004PhRvD..69l3512G,2004PhRvD..70d3527S,Keresztes:2009vc,Quiros:2009mz,Avelino:2010qn}. We apply it to the particular case of quintessence and tachyon dark energy models with a constant dark energy equation of state parameter, $w_e$, with the same background dynamics, considering both dark energy and unified dark energy roles \cite{Beca:2005gc,Avelino:2008cu} for the tachyon scalar field. We investigate the amount of fine-tuning of the corresponding scalar field potentials which would be required in order to obtain a constant  $w_e \neq -1$ around the present epoch. We also discuss the implications of our results for the reconstruction of the dark energy equation of state parameter with redshift using varying couplings.

Throughout this work we use units in which $c={4\pi G}=H_0=1$, where $c$ is the speed of light in vacuum, $G$ is the gravitational constant, $H$ is the Hubble parameter and the subscript `0' refers to the present time.

%%%%%%%%%%%%%%%%%%%%%%%%%%

\section{FRW models with a generic scalar field and matter}

We shall consider models with matter and a real scalar field, $\chi$, minimally coupled to gravity described by the action
\be\label{model}
S=\int\,d^4x\;{\sqrt{-g}\;\left(-\frac14\,R+{\mathcal L}_m+{\mathcal L_e(\chi,X)}\right)}\,,
\ee
where ${\mathcal L}_m$ and ${\mathcal L_e(\chi,X)}$ are, respectively, the matter and scalar field Lagrangians, $X=\chi_{,\mu}  \chi^{,\mu}/2$ and a comma represents a partial derivative. In the following it will be assumed that $\chi$ plays a dark energy role.

The energy-momentum tensor of the real scalar field can be written in a perfect fluid form
\begin{equation}\label{eq:fluid}
T^{\mu\nu}_{[e]}= (\rho_e+ p_e) u^\mu u^\nu - p_e g^{\mu\nu} \,,
\end{equation}
by means of the following identifications
\begin{equation}\label{eq:new_identifications}
u_\mu = \frac{\chi_{, \mu}}{\sqrt{2X}} \,,  \quad \rho_e = 2 X {\mathcal L}_{e,X} - {\mathcal L}_e \, ,\quad p_e =  {\mathcal L}_e(X,\chi)\,.
\end{equation}
In Eq.~(\ref {eq:fluid}), $u^\mu$ is the 4-velocity field describing the motion of the fluid (for timelike $\chi_{, \mu}$), while $\rho_e$ and $p_e$ are its proper energy density and pressure, respectively. The equation of state parameter, $w_e$ is equal to 
\begin{equation} 
\label{eq:w}
w_e \equiv \frac{p_e}{\rho_e} = \frac{{\mathcal L}_e}{2X {\mathcal L}_{e,X}  - {\mathcal L}_e}\,, 
\end{equation} 
and the sound speed squared is given by
\begin{equation}
\label{eq:cs2}
c_{s[e]}^2 \equiv \frac{p_{e,X}}{\rho_{e,X}}=\frac{{\mathcal L}_{e,X}}{{\mathcal L}_{e,X}+2X{\mathcal L}_{e,XX}}\,.
\end{equation}

The components of the energy-momentum tensor of the matter field are
\begin{equation}\label{eq:fluidm}
T^{\mu\nu}_{[m]}= \rho_m v^\mu v^\nu \,,
\end{equation}
where $v^\mu$ is the 4-velocity field of the matter field and $\rho_m$ is its proper energy density. Its proper pressure, $p_m$, 
is equal to zero so that both the equation of state parameter and the sound speed vanish ($w_m=p_m/\rho_m=0$ and $c_{s[m]}^2=0$).

Consider a flat Friedmann-Robertson-Walker background with line element
\be
ds^2=dt^2 - a^2(t)\left(dx^2+dy^2 +dz^2\right) \,,
\ee
where $t$ is the physical time and $x$, $y$ and $z$ are comoving spatial coordinates.
Einstein's equations then imply 
\bes
\ben
\label{eq:ein1}
H^2&=&\frac23\,\rho\,,
\\
\label{eq:ein2}
{\dot H}&=&-(\rho+p)\,,
\een
\ees
where $H={\dot a}/a$ is the Hubble parameter, $\rho=\rho_m+\rho_e$ is the total energy density, $p=p_e$ is the total pressure and a dot represents a derivative with respect to physical time.
Energy-momentum conservation for the both matter and dark energy components leads to
\bes
\ben
\label{eq:ein3}
{\dot \rho_m} = -3H\rho_m \,,
\\
\label{eq:ein4}
{\dot \rho_e} = -3H(1+w_e)\rho_e\,,
\een
\ees
thus implying that $\rho_m=\rho_{m0} a^{-3}$, $\rho_e=\rho_{e0} a^{-3(1+w_e)}$ (assuming a constant $w_e$). Hence, Eq.\eqref{eq:ein1} can also be written as
\be\label{ha}
H^2=\Omega_{e0}a^{-3(1+w_e)}+\Omega_{m0}a^{-3}\,,
\ee
where $\Omega_{m0}=2\rho_{m0}/3$, $\Omega_{e0}=2\rho_{e0}/3=1-\Omega_{m0}$. In the following we consider a class of solutions satisfying
\bes
\ben
\label{hzphi}
\frac{\dot{a}}{a}=H(\chi)\,,
\\
\label{hzphi1}
\dot\chi=Z(\chi)\,,
\een
\ees
where $H(\chi)$ and $Z(\chi)$ are, in principle, arbitrary functions of the scalar field, $\chi$.
The background dynamics fully determines the (global) equation of state parameter
\be
w=\frac{p}{\rho}=\frac{w_e}{1+\Delta a^{3w_e}}=-1 -\frac23 \frac{H_{,\chi}Z}{H^2} \,,
\ee
where $\Delta=\Omega_{m0}/\Omega_{e0}$.

Quintessence and tachyon scalar fields will be described by different greek letters ($\phi$ and $\psi$, respectively). Also, we shall employ different letters, $V$ and $U$, for the quintessence and tachyon potentials and use the notation $Z={\dot \phi}$ and ${\mathcal Z}={\dot \psi}$ in order to distinguish the time derivative of quintessence and  tachyon scalar fields, respectively.

%%%%%%%%%%%%%%%%%%%%%%%%%%%%

\section{Quintessence scalar field}

Here, we investigate a family of scalar field models described by the Lagrangian
\be
\label{td}
{\mathcal L}=\frac12 \phi_{,\mu} \phi^{,\mu}-V(\phi)\,,
\ee
where $V$ is the quintessence field potential (see also \cite{Bazeia:2005tj,Bazeia:2006mh}). The corresponding density and pressure are given by
\be
\rho_e=\frac12{Z^2} + V\,,\;\;\;\;p_e=\frac12{Z^2} - V\,,
\ee
so that
\be\label{ws}
w_e=\frac{Z^2/2 - V(\phi)}{Z^2/2 +V(\phi)}\,.
\ee
Eqs.~(\ref {eq:ein1}) and (\ref {eq:ein2}) can now be written as 
\bes\label{hs}
\be\label{hs1}
H^2=\frac13Z^2+\frac23V+\frac23Y\,,
\ee
\be\label{hs2}
H_{,\phi}Z=-Z^2-Y\,,
\ee
\ees
where $Y=\rho_m=3  \, \Omega_{m0} a^{-3}/2$. Hence, the scalar field potential becomes
\be\label{vhz}
V=\frac32H^2+H_{,\phi}Z+\frac12Z^2\,,
\ee
with the constraint given by Eq. \eqref{eq:ein3}
\be\label{vinculo1}
ZY_{,\phi}+3HY=0\,,
\ee
where
\be\label{yphi}
Y=-Z(H_{,\phi}+Z)\,.
\ee
If $\rho_{m0}=0$ then $Z=-H_{,\phi}$. In this limit one obtains the first-order formalism introduced in \cite{Bazeia:2005tj}.

%%%%%%%%%%%%%%%%%%%%%%%

\subsection{Constant $w_e$}

For a constant $w_e$, Eq. \eqref{ws} implies
\be\label{zw}
Z=\pm\left(2V\frac{1+w_e}{1-w_e}\right)^{1/2}\,.
\ee
In the following we will omit the $\pm$ sign and shall only consider the solution with $Z > 0$. However, this assumption may be relaxed since the model is invariant by the transformation $\phi \to -\phi$, $V(\phi) \to V(-\phi)$. From Eqs. \eqref{hs1},  \eqref{hs2} and \eqref{zw}, one obtains
\be
\label{h2V}
H^2=\frac2{9(1-w_e^2)}\frac{V^2_{,\phi}}{V}\,.
\ee
Multiplying both sides of Eq. \eqref{h2V} by $Z^2={\dot \phi}^2$ and using Eq.  \eqref{zw} one finds 
\be\label{va}
V=V_0\;a^{-3(1+w_e)}\,.
\ee

Using Eqs. \eqref{ha}, \eqref{zw} and \eqref{va}, taking into account that $\Delta=\Omega_{m0}/\Omega_{e0}$ and $\rho_{e0}=2V_0/(1-w_{e})$ so that $V_0=3 \, \Omega_{e0}(1-w_e)/4$, one may show that
\be\label{dphia}
\frac{d\phi}{da}=\frac{Z}{aH} = \left(\frac{3(1+w_e)/2}{\Delta \, a^{2+3w_e}+a^2}\,\right)^{1/2}\,,
\ee
whose solution is given by
\be\label{phia}
\phi=2r\ln\left(\frac{a^{3w_e/2} (1+\sqrt{1+\Delta})}{1+\sqrt{1+\Delta \, a^{3w_e}}}\right)\,,
\ee
where $r=\sqrt{3(1+w_e)/2}/(3w_e)$ and the integration constant was chosen so that $\phi_0=0$. Inverting Eq. \eqref{phia} one obtains
\be
a=\left(\frac{2(1+\sqrt{1+\Delta})\,e^{\phi/(2r)}}{2(1+\sqrt{1+\Delta})+\Delta(1-e^{\phi/r})}\right)^{2/(3w_e)}\,,
\ee
and using Eq. \eqref{phia} one obtains the potential \cite{Sahni:1999gb,2001IJMPD..10..231D}
\be
\label{vphi}
V=V_0\left(\frac{2(1+\sqrt{1+\Delta})\,e^{\phi/(2r)}}{2(1+\sqrt{1+\Delta})+\Delta(1-e^{\phi/r})}\right)^{-2(1+w_e)/w_e}\,.
\ee

The scalar field potential has the form
\be\label{ve}
V \propto e^{-\sqrt{6(1+w_e)}\phi}\,,
\ee
deep into the dark energy dominated era ($a>>1$), and 
\be\label{vm}
V \propto (\phi-\phi_{m*})^{2(1+w_e)/w_e}\,,
\ee
where $\phi_{m*}$ is a constant, deep into the matter dominated era ($a<<1$). The rapid change in the shape of the potential around the present epoch is due to the fact that, although the function $V(a)$ has the same form in the dark matter and dark energy dominated eras, the dynamics of 
quintessence field, $\phi(a)$, is significantly affected by the change in the universe dynamics around the present time. As a consequence, in order that $w_e = {\rm const} \neq -1$, the shape of the scalar field potential, $V(\phi)$, needs to compensate for that change, thus requiring a significant amount of fine-tuning.  These results are in agreement with \cite{Avelino:2009ze,2001IJMPD..10..231D}.

%%%%%%%%%%%%%%%%%%%%%%%%%%%%%

\section{Tachyon scalar field}

Now, we examine the family of scalar field models described by the Lagrangian
\be
{\mathcal {L}}=-U(\psi)\sqrt{1-\psi_{,\mu} \psi^{,\mu}}\,,
\ee
where $U$ is the potential for the tachyonic real scalar field, $\psi$. In this case, considering \eqref{eq:new_identifications}, the energy density and pressure are given by
\be
\rho=U\frac1{\sqrt{1-{\mathcal Z}^2}}\,\;\;\;\;p=-U\sqrt{1-{\mathcal Z}^2}\,,
\ee
with implies that $w_e=-1+{\mathcal Z}^2$,
and 
Eqs.~(\ref {eq:ein1}) and (\ref {eq:ein2}) can be written as 
\bes\label{ht}
\be\label{ht1}
H^2=\frac23\frac{U}{\sqrt{1-{\mathcal Z}^2}}+\frac23 Y\,,
\ee
\be\label{ht2}
H_{,\psi}{\mathcal Z}=-\frac{{\mathcal Z}^2}{\sqrt{1-{\mathcal Z}^2}}U-Y\,.
\ee
\ees
Hence, the potential is
\be\label{uhz}
U=\frac{3H^2+2H_{,\psi}{\mathcal Z}}{2\sqrt{1-{\mathcal Z}^2}}\,,
\ee
with the constraint given by Eq.\eqref{eq:ein3}
\be
\label{tachcons}
{\mathcal Z}Y_{,\psi}+3HY=0\,,
\ee
where
\be\label{ypsi}
Y=-\frac{{\mathcal Z}(2H_{,\psi}+3H^2{\mathcal Z})}{2(1-{\mathcal Z}^2)}\,.
\ee
If $\rho_{m0}=0$ then ${\mathcal Z}=-2{H_{,\psi}}/(3H^2)$, which is the case studied in \cite{Bazeia:2005tj} using a first-order formalism.

%%%%%%%%%%%%%%%%%%%%%%%%%%

\subsection{Constant $w_e$}

If we require $w_e$ to be a constant, then 
\be\label{zpsi}
{\mathcal Z}=\dot\psi=\pm \sqrt{1+w_e}\,.
\ee
In the following we omit the $\pm$ sign and shall only consider the solution with ${\mathcal Z} > 0$. However, this assumption may be relaxed since the model is invariant by the transformation $\psi \to -\psi$, $U(\psi) \to U(-\psi)$. Using Eqs. \eqref{ht1},  \eqref{ht2} and \eqref{zpsi}, one obtains
\be
H=-\frac1{3\sqrt{1+w_e}}\frac{U_{,\psi}}{U}\,,
\ee
which implies
\be\label{ua}
U=U_0\;a^{-3(1+w_e)}\,,
\ee
as in the case of a standard scalar field (see Eq.\eqref{va}). It is also possible to show, analogously to what was done for the standard scalar field, that
\be
\frac{d\psi}{da}=\frac{\mathcal Z}{aH} = \sqrt{1+w_e}\left(\frac{\Omega_{m0}}{a}+\frac{\Omega_{e0}}{a^{1+3w_e}}\right)^{-1/2}\,,
\ee
which gives
\be\label{psia}
\psi=\frac23\sqrt{\frac{1+w_e}{\Omega_{m0}}}\; \left[ a^{3/2}\,_2F_1
\left(\frac12,-\frac{1}{2w_e},1-\frac1{2w_e};-\frac{a^{-3w_e}}{\Delta}\right)-\,_2F_1
\left(\frac12,-\frac{1}{2w_e},1-\frac1{2w_e};-\frac{1}{\Delta}\right)\right]\,,
\ee
where the integration constant was chosen so that $\psi_0=0$. The duality between quintessence and tachyon models for constant $w_e$, can be written as
\be\label{psiphi}
\psi=\sqrt{1+w_e}\int{\frac{d\phi}{Z}}\,.
\ee

Analytically, the relation between the two scalar fields is non invertible. However, using Eq. \eqref{psiphi}, we may find the correspondence in the limit cases. In the dark energy dominated era ($a \gg 1$)
\be
\psi-\psi_{e*} \propto \exp\left( \frac{\sqrt{6(1+w_e)}}{2}\phi\right)\,,
\ee
where $\psi_{e*}$ is a constant. Using Eq.\eqref{ve} it is possible to find the corresponding  tachyonic potential
\be
U   \propto (\psi-\psi_{e*})^{-2}\label{uene}\,.
\ee
From Eq. \eqref{psiphi} one obtains, in the matter dominated era ($a \ll 1$), 
\be
\phi\propto(\psi-\psi_{m*})^{-w_e}\,,
\ee
where $\psi_{m*}$ is a constant. The corresponding  tachyonic potential is given by
\be
U \propto (\psi-\psi_{m*})^{-2(1+w_e)}\label{umat}\,.
\ee

\begin{figure}[ht!]
\includegraphics[{height=6cm,width=7.6cm}]{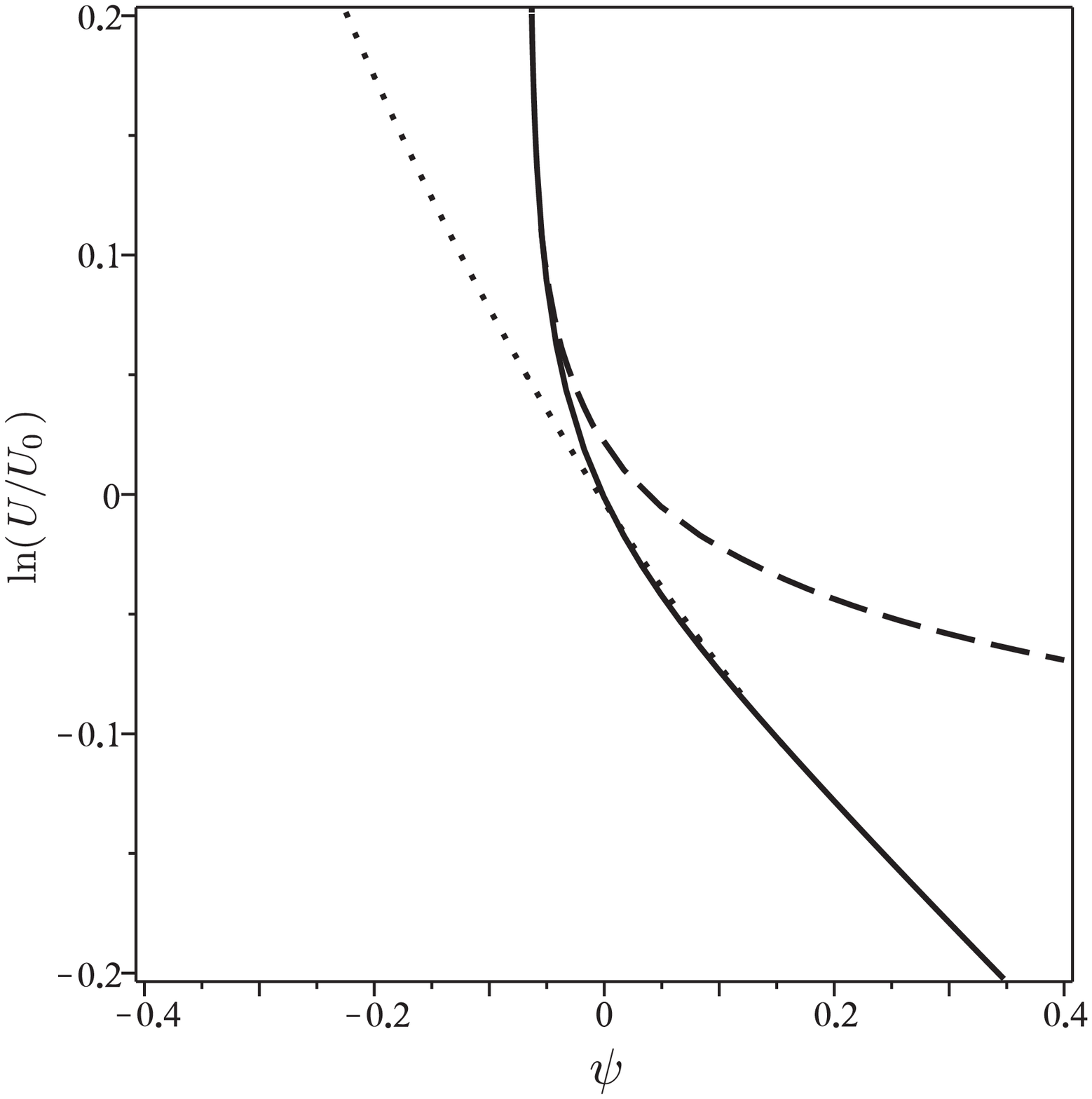}
\caption{The solution for $U(\psi)$ assuming that $w_0=-0.97$ at all times (solid line), as well as the analytical solutions for the tachyon potential, computed using Eqs. (\ref {umat}) and  (\ref {uene}), valid deep into the matter era (dashed line) and dark energy era (dotted line) respectively. The value of $\psi$ at the present time is $\psi_0=0$.}
\end{figure}

Fig.~1 shows the solution for $U(\psi)$ assuming that $w_e=-0.97$ at all times (solid line), as well as the analytical solutions, computed using Eqs. \eqref{umat} or  \eqref{uene}, valid deep into the matter and dark energy eras (dashed and dotted lines, respectively). The initial conditions for the constant $w_e$ solution were chosen so that $\psi_0=0$ and the constants $\psi_{e*}$ and $\psi_{m*}$ were determined by requiring that the analytical solutions computed using Eqs. (\ref {umat}) or  (\ref {uene}) fitted the constant $w_e$ results obtained deep into the matter and dark energy dominated eras, respectively. In this paper we take $\Omega_{m0}=0.27$ and $\Omega_{e 0}=0.73$ as  favored by the seven-year WMAP results \cite{Komatsu:2010fb}. Fig.~1 shows that, in order that $w_e={\rm const}$, the shape of the potential must be fine-tuned around $\psi=\psi_0=0$. Otherwise,  the equation of state parameter would change rapidly around the present time. 

\begin{figure}[ht!]
\includegraphics[{height=6cm,width=7.6cm}]{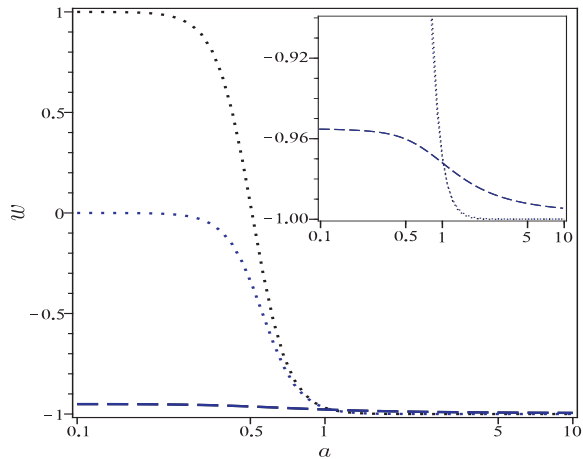}
\caption{The evolution of the equation of state parameter computed with the constant $w_e$ matter era quintessence/tachyon potentials given by Eqs. (\ref {vm}) and (\ref{umat}) (dashed lines) or with the constant $w_e$ dark energy era quintessence/tachyon potentials given by Eqs. (\ref {ve}) and (\ref {uene}) (dotted lines). Significant differences between the results obtained for the quintessence and tachyon fields only appear for $w_e$ significantly larger than $-1$.}
\end{figure}

This is also shown in Fig.~2, where we plot the evolution of the equation of state parameter with the potentials $V$ and $U$ designed to produce a constant $w_e$ deep into the matter and dark energy dominated eras (dashed and dotted lines, respectively). As expected the figure shows that $w_e$ is roughly constant deep inside the matter era (dashed line) or deep inside the dark energy era (dotted line) but there is a rapid change in $w_e$ in the transition between them, with $|w_{e0}-w_e(z=1)|/|w_{e0}+1| \gsim 1$ (here $z=1/a-1$ is the redshift). In fact, the evolution of the equation of state parameter computed with the constant $w_e \neq -1$ dark energy era quintessence potential is not consistent with observations, since the scalar field would completely dominate the energy density of the universe at moderate redshifts, when $w_e$ becomes close to unity. This is no longer necessarily true for the tachyon field since, in this case, the equation of state parameter cannot be larger than zero. On the other hand, the evolution of the equation of state parameter computed with the constant $w_e=-0.97$ matter era quintessence/tackyon potentials is in agreement with observations (the equation of state parameter of the dark energy is always smaller than $-0.95$).

The cosmology obtained considering a tachyon model for dark energy is equivalent to a standard quintessence cosmology up to first order in $Z/V$ (or equivalently ${\mathcal Z}$). Hence, for slow rolling fields with $w_e \sim -1$ there is a simple correspondence between the background evolution predicted in both models, even if $w_e$ is not a constant, corresponding to $V=U$ and $\phi=\psi {\sqrt U}$. This is the reason for the similarity between the results presented in Fig. 2 for the tackyon (+ dark matter) and quintessence models with $w_e \sim -1$ (see the inset of Fig.~2). In fact, a similar result is to be expected, in the slow rolling limit, in the case of a generic Lagrangian admitting  an expansion of the form
\begin{equation}\label{lagran}
{\mathcal L}=-V(\chi)+f(\chi)X+g(\chi) X^2+...\,,
\end{equation}
where $f$ and $g$ are functions of a scalar field $\chi$.
Significant differences between the quintessence and tackyon models only appear for $w_e$ significantly larger than $-1$. In particular, the equation of state parameter for the tachyon field can never become greater than zero, while the equation of state parameter of the quintessence field may vary in the interval $[-1,1]$.

\subsection{Unified dark energy}

The tachyon has also been proposed as a unified dark energy candidate. In fact, it is possible to show that there is a duality, at the background level, between pure tachyon models described by a scalar field $\psi$ and quintessence models with dark energy, described by a scalar field $\phi$, and dark matter. In that case the correspondence between the tachyon and quintessence scalar fields is given by
\be\label{psiphi2}
\psi=\pm\sqrt{\frac23}\int{\left(-\frac{H_{,\phi}}{Z}\right)^{1/2}\frac{d\phi}{H}}\,.
\ee
In the following we omit the $\pm$ sign and shall only consider the solution with ${\mathcal Z} > 0$.
The corresponding tachyonic potential  can be written  as
\be\label{uphi}
U=\frac32 H^2\left(1+\frac23\frac{Z H_{,\phi}}{H^2}\right)^{1/2}\,.
\ee
The evolution of $\psi$ with the scale factor is given by
\be
\psi(a)=\frac1{\sqrt{\Omega_{e0}}} \int{\frac{\left(\Delta a^{3w_e}+(1+w_e)\right)^{1/2}}{\Delta a^{3w_e}+1}a^{(1+3w_e)/2}da}\label{psiude}\,,
\ee
and
\be
U(a)=\frac32\sqrt{-w_e\Omega_{e0}}\left(\Omega_{m0}a^{-3(2+w_e)}+\Omega_{e0}a^{-6(1+w_e)}\right)^{1/2}\,.
\ee
If $w_e=-1$ then Eq. \eqref{psiude} gives 
\be
\psi-\psi_*=\frac{2}{3 {\sqrt {\Omega_{e0}}}} \arctan \left[ {\sqrt{\frac{a^3}{\Delta}}} \right]\,,
\ee
with $\psi_*=-2  \arctan (\Delta^{-1/2})/(3 {\sqrt {\Omega_{e0}}})$. This in turn implies 
that 
\be
U(\psi)=\frac{3 \,  \Omega_{e0}}{2 \left| \sin \theta \right|}\,,
\ee
with $\theta=3(\psi-\psi_*) {\sqrt {\Omega_{e0}}}/2$. As $\theta \to \pi/2$ (when $a \to \infty$) the tachyon potential $U$ tends to  
the constant $3 \, \Omega_{e0}/2$. On the other hand, for $a \ll 1$ (for $\theta \sim 0$ and $\psi \sim \psi_*$)  the tachyon potential 
$U$ is roughly proportional to $(\psi-\psi_*)^{-1}$. Hence, if the tachyon field plays the role of both dark matter and dark energy then the shape of the tachyon potential $U$ has to be fine tuned (even assuming that $w_e=-1$).

\section{Varying couplings}

We now consider the possibility that the dark energy scalar field is also responsible for the cosmological variation of fundamental couplings, such as the fine structure constant, $\alpha$ (or the proton-to-electron mass ratio $\mu=m_p/m_e$). It has been demonstrated \cite{Copeland:2003cv,Nunes:2003ff,Avelino:2006gc,Avelino:2009fd} that the reconstruction of the evolution of the equation of state parameter of dark energy would be possible using observations of the evolution of $\alpha$ with redshift, assuming that the dark energy is described by a standard scalar field. If the fine structure constant, $\alpha$, is a linear function of $\phi$ then one has
\be
\label{linear}
\frac{\Delta \alpha}{\alpha}=\beta \Delta \phi\,,
\ee
with $\Delta \alpha=\alpha-\alpha_0$, $\Delta \phi=\phi-\phi_0$ and $\beta={\rm const}$. This is no longer the case if one of these assumptions is relaxed. For example, if dark energy is described by a tachyon field and $X={\rm const}$ then $w_e={\rm const}$. However, if we attempted to reconstruct evolution of $w_e$ (wrongly) assuming a standard scalar field one would obtain
\be
w_e(a)=\frac{w_{e0}+3(1+w_{e0})\ln a}{1-3(1+w_{e0})\ln a}\,.
\ee
This confirms that the success of the dark energy reconstruction using varying couplings is crucially dependent on the properties of the scalar field lagrangian \cite{Avelino:2008dc}, even if the (very strong) assumption given by  Eq. \eqref{linear} turns out to be valid.

In the unified scenario the problem is even worse. If $w_e=-1$ then $Z=0$ or ${\mathcal Z}=0$ in the standard quintessence or tachyon (+ dark matter) scenarios, respectively. However, in the unified dark energy scenario this is no longer the case since, although the equation of state parameter of the tachyon field must be very close to $-1$ at late times (${\mathcal Z} \sim 0$), at early times, deep in the matter era, the equation of state parameter must be very close to zero (${\mathcal Z} \sim 1$). This poses a fundamental problem for the reconstruction of the dark energy equation of state using varying couplings.

\section{Ending Comments} 

In this paper we have further explored the correspondence between quintessence and tachyon models, giving particular attention to dark energy models with a constant dark energy equation of state parameter, $w_e$.  It was shown that a large fine-tuning of 
the potentials is required in order to obtain $w_e \sim {\rm const} \neq -1$ around the present epoch in all models investigated. 
This result is a consequence of the dramatic change in the background evolution in the transition between the matter and 
dark energy dominated epochs, which must be compensated by a fine-tuning of the dark energy model. We have 
demonstrated this for the special case of quintessence and tachyon dark energy models but we expect that similar results 
would hold in any dynamical dark energy model where a nearly homogeneous dark energy component is described by a scalar, vector or tensor field. We have also demonstrated that the evolution of the scalar fields can be quite different in dual (at the background level) quintessence and tachyon models and we have shown that this may be a serious drawback for the proposed reconstruction of the evolution of the dark energy equation of state with redshift using varying couplings. \\

The authors would like to thank Alexandre Barreira for useful comments and CAPES, CNPq, Brasil and FCT, Portugal for partial support.

\bibliography{ALR}

\end{document}